\documentclass[
    ,final            % use final for the camera ready runs
  ]
  {aipproc}

\layoutstyle{6x9}

\begin{document}

\title{The CLEO-c Research Program}

\author{Holger St\"ock}{
  address={University of Florida, Department of Physics, PO Box 118440,
           Gainesville, FL 32611-8440, USA}
}

\begin{abstract}

In spring 2003, the B physics era ended for the CLEO experiment with a final run
at the $\Upsilon(5S)$ resonance. Over the summer the experiment and the CESR 
accelerator
were modified to operate at lower center-of-mass energies between 3 and 5 GeV.
In September 2003 the CLEO-c detector
has begun to take its first data at the $\psi(3770)$
resonance, with which a new era for the exploration of the charmonium sector 
begins. 
The CLEO-c research program presented here will include studies of leptonic,
semileptonic
and hadronic charm decays, searches for exotic, gluonic matter and test for
new physics beyond the Standard Model. 

\end{abstract}

\maketitle

%%%%%%%%%%%%%%%%%%%%%%%%%%%%%%%%%%%%%%%%%%%%
%% MAINMATTER
%%%%%%%%%%%%%%%%%%%%%%%%%%%%%%%%%%%%%%%%%%%%

\section{Introduction}

The CLEO-c physics program includes a variety of measurements
that will contribute to the understanding of important Standard Model processes
as well as provide the opportunity to probe the physics that lies beyond the
Standard Model. The dominant themes of this program are measurement of absolute
branching ratios for charm mesons with the precision of the order of 1 - 2\%
(depending upon the mode), determination of charm meson decay constants and of
the CKM matrix elements $\mid V_{cs} \mid$ 
and $\mid V_{cd} \mid$ at the 1 - 2\% level and
investigation of processes in charm and $\tau$ decays, that are expected to be
highly suppressed within the Standard Model. Hence, a reconfigured CESR
electron-positron collider operating at a center of mass energy range between
3 and 5 GeV together with the CLEO detector will give significant
contributions to our understanding of fundamental Standard Model properties.

\section{Run Plan and Data Sets}

From the year 2003 to 2006 the CESR accelerator will be operated at
center-of-mass energies corresponding to $\sqrt{s} \sim 4140 MeV$, $\sqrt{s}
\sim 3770 MeV$ ($\psi$'') and $\sqrt{s} \sim 3100 MeV$ ($J/\psi$). Taking
into account the anticipated luminosity which will range from $5 \times 10^{32}
cm^{-2} s^{-1}$ down to about $1 \times 10^{32} cm^{-2} s^{-1}$ over this
energy range, the run plan will yield $3 fb^{-1}$ each at the $\psi$'' and at
$\sqrt{s} \sim 4140 MeV$ above $D_s \bar{D_s}$ threshold and $1 fb^{-1}$ at the
$J/\psi$. These integrated luminosities correspond to samples of 1.5 million
$D_s \bar{D_s}$ pairs, 30 million $D \bar{D}$ pairs and one billion $J/\psi$
decays. As a point of reference, these datasets will exceed those of the
Mark III experiment by factors of 480, 310 and 170, respectively.
Table \ref{tab:runplan} summarizes the run plan.
\newpage

\begin{table}
\begin{tabular}{llll}
\hline
    \tablehead{1}{l}{b}{Year}
  & \tablehead{1}{l}{b}{Resonance}
  & \tablehead{1}{l}{b}{Anticipated\\Luminosity ($fb^{-1}$)}
  & \tablehead{1}{l}{b}{Reconstructed Events}   \\
\hline
1 & $\psi(3770)$             & $\sim$ 3 & 30M $D\bar{D}$ , 6M tagged $D$ \\
2 & $\sqrt{s} \sim 4140 MeV$ & $\sim$ 3 & 1.5M $D_s\bar{D_s}$ , 0.3M tagged $D_s$  \\
3 & $\psi(3100)$             & $\sim$ 1 & 60M radiative $J/\psi$  \\
\hline
\end{tabular}
\caption{The 3-year CLEO-c run plan}
\label{tab:runplan}
\end{table}

\par
In addition, prior to the conversion to low energy a total amount of $4
fb^{-1}$ spread over the $\Upsilon(1S)$, $\Upsilon(2S)$, $\Upsilon(3S)$
and $\Upsilon(5S)$
resonances is taken to launch the QCD part of the program. These data sets will
increase the available $b\bar{b}$ bound state data by more than an order of
magnitude.

\section{Hardware Requirements}
The conversion of the CESR accelerator for low energy operation requires
the addition of 18 meters of wiggler magnets to enhance transverse cooling of
the beam at low energies. 6 of 14 wigglers were installed in summer 2003 with
additional 6 wigglers scheduled for installation in 2004. 
In the CLEO III detector the silicon vertex detector was replaced by a
small, low mass inner drift chamber.
In addition, the solenoidal field will be
reduced to 1.0 T. No other requirements are necessary.

\section{Physics Program}

The following sections will outline the CLEO-c physics program.
The first section will focus on the Ypsilon spectroscopy, the second section
will describe the charm decay program, the third section
will give an overview about the exotic, gluonic matter studies and the last
section will descibe the oportunities for probing of new
physics beyond the Standard Model.

\subsection{Ypsilon Spectroscopy \label{sec:ypsilon}}

From fall 2001 to spring 2003 CLEO has collected $4 fb^{-1}$ of
data on the $\Upsilon$
resonances below the $\Upsilon(4S)$, as well as at the $\Upsilon(5S)$
resonance, which is currently beeing analyzed. 
\par
\begin{figure}[ht]
  \includegraphics[height=.4\textheight]{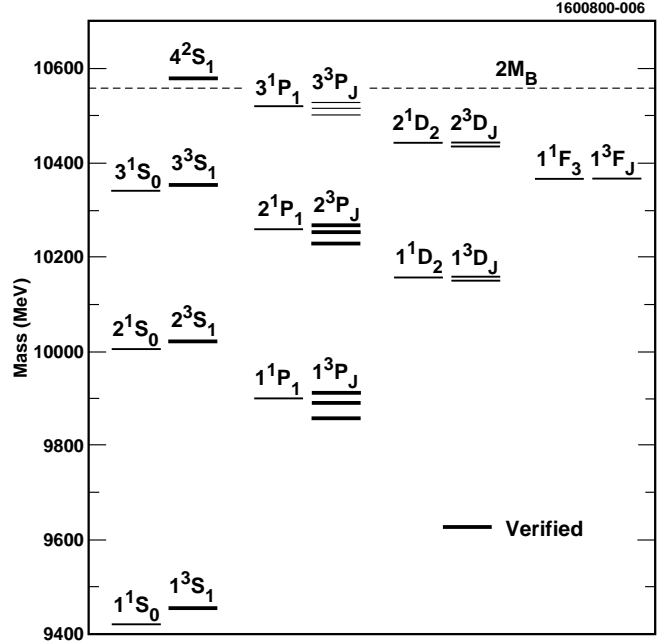}
  \caption{Approximate levels of the $b\bar{b}$ states. The name associated with
the spin-parity assignments are $^1S_0 = \eta_b$, $^3S_1 = \Upsilon$,
$^1P_1 = h_b$ and $^3P_J = \chi_b$ (triplets with J = 0,1,2).}
  \label{fig:Ystates}
\end{figure}

\noindent
So far, the only established states below $B\bar{B}$ threshold are the three
vector singlet $\Upsilon$ resonances ($^3S_1$) and the six $\chi_b$ and
$\chi'_b$ (two triplets of $^3P_J$) that are accessible from these parent
vectors via E1 radiative transitions (see Figure \ref{fig:Ystates}). 
By collecting substantial data samples
at the $\Upsilon(1S)$, $\Upsilon(2S)$ and $\Upsilon(3S)$, CLEO will address a
variety of outstanding physics issues.

\begin{itemize}

\item Discovery of $\eta_b$ and Observation of $h_b$ \newline
The $\eta_b$ is the ground state of $b\bar{b}$. Most present theories 
\cite{bib:EFI01-10} indicate
the best approach would be the hindered M1 transition from the $\Upsilon(3S)$,
with which CLEO might have a signal of $5 \sigma$ significance in $1 fb^{-1}$
of data. In the case of the $h_b$, CLEO established an upper limit of
${\cal B}(\Upsilon(3S) \rightarrow \pi^+ \pi^- h_b)$ $<$ 0.18\% at 90\%
confidence level \cite{bib:hb}. This result, based on $\sim 110 pb^{-1}$, 
already
tests the theoretical predictions \cite{bib:hb2} for this transition which range
from 0.1 - 1.0\%. The resonance run program will measure the mass of the $h_b$,
assuming the predictions are valid, to $\sim 5 MeV$.

\item Observation of $1^3D_J$ states \newline
The $b\bar{b}$ system is unique as it has states with L = 2 that lie below
the open-flavor threshold. These states have been of considerable theoretical
interest, as indicated by many predictions of the center-of-gravity of the
triplet and by a recent review \cite{bib:EFI01-14}. 
In an analysis of the $\Upsilon(3S)$ CLEO data sample the 
$\Upsilon(1^3D_2)$ state could already be observed in the
four-photon cascade
$\Upsilon(3S) \rightarrow \gamma_1\chi'_b
\rightarrow \gamma_1\gamma_2 \Upsilon(^3D_J) \rightarrow \gamma_1\gamma_2\gamma_3
\chi_b \rightarrow \gamma_1\gamma_2\gamma_3\gamma_4\ell^+\ell^-$.
The mass of the $\Upsilon(1^3D_2)$ state is determined to
$10161.1 \pm 0.6 \pm 1.6 MeV/c^2$ \cite{bib:lp2003}.

\item Search for glueball candidates in radiative $\Upsilon(1S)$ decays \newline
The BES collaboration has reported signals for a glueball candidate
\cite{bib:radiative} in radiative $J/\psi$ decay - a glue-rich environment. 
Naively
one would expect the exclusive radiative decay to be suppressed in $\Upsilon$
decay by a factor of roughly 40, which implies product branching fractions for
$\Upsilon$ radiative decay of $\sim 10^{-6}$. With $1 fb^{-1}$ of data and
efficiencies of around 30\% one can expect $\sim$ 10 events in each of the
exclusive channels, which would be an important confirmation of the $J/\psi$
studies.

\end{itemize}

\subsection{Charm Decays \label{sec:charm}}
The observable properties of the charm mesons are determined by the strong
and weak interactions. As a result, charm mesons can be used as a laboratory
for the studies of these two fundamental forces. Threshold charm experiments
permit a series of measurements that enable direct study of the weak
interactions of the charm quark, as well as tests of our theoretical technology
for handling the strong interactions.

\subsubsection{Leptonic Charm Decays}
Measurements of leptonic decays in CLEO-c will benefit from the use of fully 
tagged $D^+$ and $D_s$ decays available at the $\psi(3770)$ and at 
$\sqrt{s} \sim 4140 MeV$. The leptonic decays $D_s \rightarrow \mu\nu$
are detected in tagged events by observing a single charged track of the
correct sign, missing energy, and a complete accounting of the residual
energy in the calorimeter. The clear definition of the initial state, the
cleanliness of the tag reconstruction, and the absence of additional 
fragmentation tracks make this measurement straightforward and essentially
background-free. This will enable measurements of the yet barely known
leptonic decay rates for $D$ and $D_s$ to a precision of 3 - 4\% and will allow
for incisive checks of theoretical calculations of the decay constants
$f_{D}$ and $f_{D_s}$ at the 1 - 2 \%. Table \ref{tab:decayconsts} summarizes
the expected precision in the decay constant measurements.

\begin{table}[ht]
\begin{tabular}{lccc}
\hline
    \tablehead{1}{l}{b}{}
  & \tablehead{1}{l}{b}{}
  & \tablehead{2}{c}{b}{Decay Constant Error \%} \\
    \tablehead{1}{l}{b}{Decay Mode}
  & \tablehead{1}{l}{b}{Decay Constant}
  & \tablehead{1}{l}{b}{PDG 2000}
  & \tablehead{1}{l}{b}{CLEO-c}   \\
\hline
$D^+ \rightarrow \mu^+ \nu$     & $f_D$     & Upper Limit  & 2.3  \\
$D^+_s \rightarrow \mu^+ \nu$   & $f_{D_s}$ & 17           & 1.7  \\
$D^+_s \rightarrow \tau^+ \nu$  & $f_{D_s}$ & 33           & 1.6  \\
\hline
\end{tabular}
\caption{Expected decay constants errors for leptonic decay modes}
\label{tab:decayconsts}
\end{table}

\subsubsection{Semileptonic Charm Decays}
The CLEO-c program will provide a large set of precision measurements in the
charm sector against which the theoretical tools needed to extract CKM matrix
information precisely from heavy quark decay measurements will be tested and
honed.
\par
CLEO-c will measure the branching ratios of many exclusive semileptonic modes,
including
$D^0 \rightarrow K^- e^+ \nu$,
$D^0 \rightarrow \pi^- e^+ \nu$,
$D^0 \rightarrow K^{-} e^+ \nu$,
$D^+ \rightarrow \bar{K}^{0} e^+ \nu$,
$D^+ \rightarrow \pi^0 e^+ \nu$,
$D^+ \rightarrow \bar{K}^{0*} e^+ \nu$,
$D^+_s \rightarrow \phi e^+ \nu$ and
$D^+_s \rightarrow \bar{K}^{0*} e^+ \nu$.
The measurement in each case is based on the use of tagged events where the
cleanliness of the environment provides nearly background-free signal samples,
and will lead to the determination of the CKM matrix elements
$\mid V_{cs} \mid$ and $\mid V_{cd} \mid$ with a precision level of
1.6\% and 1.7\%, respectively. Measurements of the vector and axial vector form
factors $V(q^2)$, $A_1(q^2)$ and $A_2(q^2)$ will also be possible at the
$\sim$ 5\% level. Table \ref{tab:semileptonic} summarizes the proposed branching
fractional errors.

\begin{table}[ht]
\begin{tabular}{lcc}
\hline
    \tablehead{1}{l}{b}{}
  & \tablehead{2}{c}{b}{BR fractional error \%} \\
    \tablehead{1}{l}{b}{Decay Mode}
  & \tablehead{1}{l}{b}{PDG 2000}
  & \tablehead{1}{l}{b}{CLEO-c}   \\
\hline
$D^0 \rightarrow K \ell \nu$    & 5        & 1.6                       \\
$D^0 \rightarrow \pi \ell \nu$  & 16       & 1.7                       \\
$D^+ \rightarrow \pi \ell \nu$  & 48       & 1.8                       \\
$D_s \rightarrow \phi \ell \nu$ & 25       & 2.8                       \\
\hline
\end{tabular}
\caption{Expected branching fractional errors for semileptonic decay modes}
\label{tab:semileptonic}
\end{table}

HQET provides a successful description of the lifetimes of charm hadrons and
of the absolute semileptonic branching ratios of the $D^0$ and $D_s$
\cite{bib:bcp3}. Isospin invariances of the strong forces lead to corrections of
$\Gamma_{SL}(D^0) \simeq \Gamma_{SL}(D^+)$ in the order of
${\cal O}(tan^2 \Theta_C) \simeq 0.05$. Likewise, $SU(3)_{Fl}$ symmetry
relates $\Gamma_{SL}(D^0)$ and $\Gamma_{SL}(D^+)$, but a priori would
allow them to differ by as much as 30\%. However, HQET suggests that they
should agree to within a few percent. A charm factory is the best place to
measure absolute inclusive semileptonic charm branching ratios, in particular
${\cal B}(D_s \rightarrow X \ell\nu)$ and thus $\Gamma_{SL}(D_s)$.

\subsubsection{Implications of the Leptonic and Semileptonic Measurements for 
               CKM}
Every weak decay involving leptons depends on both CKM elements and on
hadronic matrix elements. As described in the sections above, CLEO-c data can 
be used for ca\-li\-brating the theoretical tools that will determine the
hadronic terms and for extracting the essential CKM elements. 
\par
Combining the leptonic and semileptonic measurements leads to ``direct''
determinations of the CKM elements $\mid V_{cd} \mid$ and 
$\mid V_{cs} \mid$. The results are
shown in Table \ref{tab:VcdVcs}. For this table LatticeQCD is assumed
and validated across a wide range of charm and onium decay measurements,
to which CLEO-c will provide decay constants with 1\% accuracy.

\begin{table}[ht]
\begin{tabular}{lcc}
\hline
    \tablehead{1}{l}{b}{Decay Mode}
  & \tablehead{1}{c}{b}{CKM Element}
  & \tablehead{1}{c}{b}{CKM Precision}   \\
\hline
$D_s \rightarrow \mu^+ \nu$     & $\mid V_{cs} \mid$ & 1.7  \\
$D_s \rightarrow \tau^+ \nu$    & $\mid V_{cs} \mid$ & 1.6  \\
$D^0 \rightarrow K^- e^+ \nu$   & $\mid V_{cs} \mid$ & 1.6  \\
\hline
$D^+ \rightarrow \mu^+ \nu$     & $\mid V_{cd} \mid$ & 2.3  \\
$D^0 \rightarrow \pi^- e^+ \nu$ & $\mid V_{cd} \mid$ & 1.7  \\
\hline
\end{tabular}
\caption{Collected results for $\mid V_{cd} \mid$ and $\mid V_{cs} \mid$}
\label{tab:VcdVcs}
\end{table}

The impact of the entire suite of CLEO-c measurements on the current
knowledge of the CKM matrix is summarized in the following paragraphs.
Just to remind the reader, the CLEO-c program of leptonic and semileptonic
measurements has two components: one of calibrating and validating
theoretical methods for calculating hadronic matrix elements, which can then be
applied to all problems in CKM extraction in heavy quark physics; and one of
extracting CKM elements directly from the CLEO-c data. The direct results
of CLEO-c are the precise determination of $\mid V_{cd} \mid$, 
$\mid V_{cs} \mid$, $f_D$,
$f_{D_s}$, and the semileptonic form factors. The precision knowledge
of the decay constants $f_D$ and $f_{D_s}$, together with the rigorous
calibration of theoretical techniques for calculating heavy-to-light
semileptonic form factors, are required for the direct extraction of CKM 
elements from CLEO-c. This also drives
the indirect results, namely the precision 
extraction
of CKM elements from experimental measurements of the $B_d$ mixing 
frequency, the $B_s$ mixing frequency, and the $B \rightarrow \pi\ell\nu$
decay rate measurements which will be done by a combination of efforts spread
across BaBar, Belle, CDF, D0, BTeV, LHCb, ATLAS and CMS.
\par
In Table \ref{tab:newCKM} the combined projections are presented. In the
determination of the CKM elements $\mid V_{cd} \mid$ 
and $\mid V_{cs} \mid$ from $B$ and $B_s$
mixing $\mid V_{tb} \mid = 1$ is used. The tabulation also includes improvement
in the direct measurement of $\mid V_{tb} \mid$ 
itself which is expected from the
Tevatron experiments \cite{bib:tevatron1}. %,bib:tevatron2}.

\begin{table}[ht]
\begin{tabular}{lll}
\hline
    \tablehead{3}{c}{b}{Present Knowledge} \\
\hline
$\delta V_{ud}/V_{ud}=$ 0.1\% & $\delta V_{us}/V_{us}=$ 1\%   & 
$\delta V_{ub}/V_{ub}=$ 25\%  \\ 
$\delta V_{cd}/V_{cd}=$ 7\%   & $\delta V_{cs}/V_{cs}=$ 16\%  & 
$\delta V_{cb}/V_{cb}=$ 5\%   \\
$\delta V_{td}/V_{td}=$ 36\%  & $\delta V_{ts}/V_{ts}=$ 39\%  & 
$\delta V_{tb}/V_{tb}=$ 29\%  \\ 
\hline
   \tablehead{3}{c}{b}{After CLEO-c}   \\
\hline
$\delta V_{ud}/V_{ud}=$ 0.1\% &
$\delta V_{us}/V_{us}=$ 1\%   & $\delta V_{ub}/V_{ub}=$ 5\%   \\
$\delta V_{cd}/V_{cd}=$ 1\%   &
$\delta V_{cs}/V_{cs}=$ 1\%   & $\delta V_{cb}/V_{cb}=$ 3\%   \\
$\delta V_{td}/V_{td}=$ 5\%   &
$\delta V_{ts}/V_{ts}=$ 5\%   & $\delta V_{tb}/V_{tb}=$ 15\%  \\
\hline
\end{tabular}
\caption{Combined projections for CKM elements at present and after CLEO-c}
\label{tab:newCKM}
\end{table}

\subsubsection{Hadronic Charm Decays}
The $D^0$ is the best known of all the charm hadrons. The CLEO and ALEPH
experiments by far provide the most precise measurements for the decay
$D^0 \rightarrow K^- \pi^+$. They use the same technique by looking at
$D^{*+} \rightarrow \pi^+ D^0$ decays and taking the ratio of the $D^0$ decays
into $K^- \pi^+$ to the number of decays with only the $\pi^+$ from the $D^{*+}$
decay detected. The dominant systematic uncertainty is the background level in
the latter sample. In both experiments, the systematic errors exceed the
statistical errors. By using $D^0\bar{D}^0$ decays, and tagging both $D$ mesons,
the background can be reduced to almost zero and the branching ratio fractional
error can be improved significantly (see Table \ref{tab:hadronic}).
\par
The $D^+$ absolute branching ratios are determined by using fully reconstructed
$D^{*+}$ decays, comparing $\pi^0 D^+$ with $\pi^+ D^0$ and using isotropic
spin symmetry. Hence, this rate cannot be determined any better than the
absolute $D^0$ decay rate using this technique. By using $D^+D^-$ decays and
a double tag technique the background can be reduced again to almost zero
which leads to a significant improvement of the branching ratio fractional
error (see Table \ref{tab:hadronic}).

\begin{table}[ht]
\begin{tabular}{lcc}
\hline
    \tablehead{1}{l}{b}{}
  & \tablehead{2}{c}{b}{BR fractional error \%} \\
    \tablehead{1}{l}{b}{Decay Mode}
  & \tablehead{1}{l}{b}{PDG 2000}
  & \tablehead{1}{l}{b}{CLEO-c}   \\
\hline
$D^0 \rightarrow K \pi$    & 2.4      & 0.5                       \\
$D^+ \rightarrow K K \pi$  & 7.2      & 1.5                       \\
$D_s \rightarrow \phi \pi$ & 25       & 1.9                       \\
\hline
\end{tabular}
\caption{Expected branching fractional errors for hadronic decay modes}
\label{tab:hadronic}
\end{table}

\subsection{Exotic, Gluonic Matter \label{sec:glue}}
\begin{figure}[ht]
  \includegraphics[height=.4\textheight]{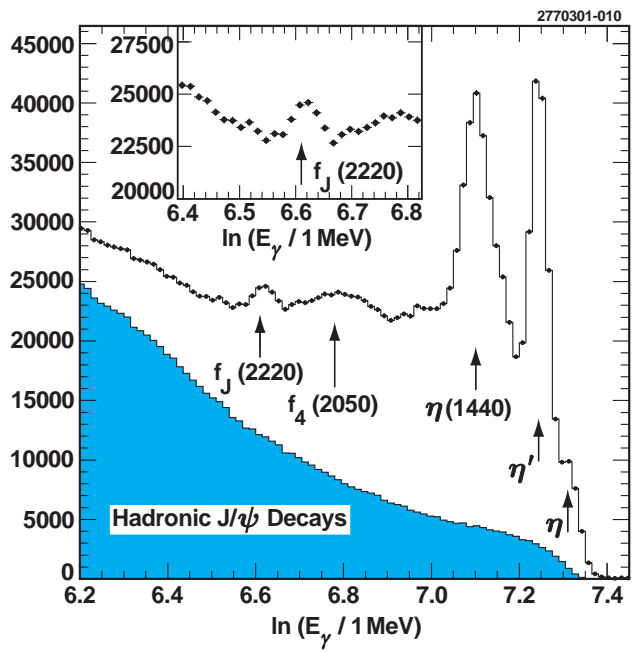}
  \caption{The inclusive photon spectrum from $J/\psi$ decays from a 
Monte Carlo simulation in CLEO-c. Signals from $\eta'$, $\eta(1440)$ and
$f_J(2220)$ are clearly visible. 
A broad signal from $f_4(2050)$ production is also evident.}
  \label{fig:inclphoton}
\end{figure}

With approximately one billion $J/\psi$ produced, CLEO-c will be the natural
glue factory to search for glueballs and other glue-rich states using
$J/\psi \rightarrow gg \rightarrow \gamma X$ decays. The region of
$1 < M_X < 3 GeV/c^2$ will be explored with partial wave analyses for
evidence of scalar or tensor glueballs, glueball-$q\bar{q}$ mixtures, exotic
quantum numbers, quark-glue hybrids and other new forms of matter predicted by
QCD. This includes the establishment of masses, widths, spin-parity quantum
numbers, decay modes and production mechanisms for any identified states, an in
detail exploration of reported glueball candidates such as the tensor candidate
$f_J(2220)$ and the scalar states $f_0(1370)$, $f_0(1500)$ and $f_0(1710)$, and
the examination of the inclusive photon spectrum $J/\psi \rightarrow \gamma$X
with $<$ 20 MeV photon resolution and identification of states with up to 100
MeV width and inclusive branching ratios above $1 \times 10^{-4}$. 
A Monte Carlo study of inclusive radiative $J/\psi$ decays in CLEO-c is
shown in Figure \ref{fig:inclphoton} based on a sample of 60 million
$J/\psi$ decays and assuming 
${\cal B}(J/\psi \rightarrow \gamma f_J(2220)) = 8 \times 10^{-4}$.
A monochromatic photon line from the $J/\psi \rightarrow \gamma f_J(2220)$
decay is clearly seen. The signal efficiency is 24\%. With 
$10^9$ $J/\psi$ decays, CLEO-c will be able to discover any narrow
resonance produced in radiative $J/\psi$ decays with inclusive branching
fractions of order $10^{-4}$ or greater.

In addition,
spectroscopic searches for new states of the $b\bar{b}$ system and for exotic
hybrid states such as $cg\bar{c}$ will be made using the $4 fb^{-1}$
$\Upsilon(1S)$, $\Upsilon(2S)$, $\Upsilon(3S)$ and $\Upsilon(5S)$
data sets. Analysis of
$\Upsilon(1S) \rightarrow \gamma X$ will play an important role in verifying
any glueball candidates found in the $J/\psi$ data.

\subsection{Charm Beyond the Standard Model}
CLEO-c will have the opportunity to probe for new physics beyond the Standard
Model. Three highlights - $D\bar{D}$-mixing, $CP$ 
violation and rare charm decays
- are discussed in the following sections.

\subsubsection{$D\bar{D}$-Mixing}
Within the Standard Model (SM), the processes which mediate the decays of
charmed quarks and antiquarks can change the ``charm'' quantum number by one
unit, $\Delta C = 1$. On the other hand, the mixing of $D^0$ and $\bar{D^0}$
necessitates changing a charm quark into an anti-charm quark, i.e. the
``charm'' quantum number must change by two units $\Delta C = 2$. This can be
arranged in the SM only at one loop level and, therefore, is naturally
suppressed. However, new physics (beyond the SM) contributions
can generate $\Delta C = 2$ interactions as well. It is for this reason
that neutral meson-antimeson mixing can provide important information
about both the SM and new physics beyond the SM. The 
$D^0 - \bar{D^0}$ system is particularly interesting in this respect as it
is the only system that is sensitive to the dynamics of the bottom-type quarks.
\par
CLEO-c will have the important experimental advantage of operating at the
$D^0 \bar{D^0}$ threshold, where the $D^0$ and $\bar{D^0}$ are produced in the
state that is quantum mechanically coherent. This allows new and simple methods
to be used to measure the $D^0 - \bar{D^0}$ mixing parameters 
\cite{bib:ddmixing1,bib:ddmixing2}. 
Thus, in addition to the ``standard'' methods
of searches for $D^0 - \bar{D^0}$ mixing, new tools and methods, unique
to running at the $\psi(3770)$ (and higher resonances), become available
for studies of these important parameters.

\subsubsection{$CP$ Violation}
In addition to indirect $CP$ violation, both SM and new physics effects can 
induce different contributions to the decay amplitudes of $D$ mesons.
This phenomenon can be traced back to the appearance of complex-valued
couplings (CKM parameters) in the $\Delta C = 1$ Lagrangian that mediates $D$
decays and leads to a $CP$-violating difference between decay rates of 
$CP$-conjugated states.
\par
The production process
\begin{displaymath}
e^+ e^- \rightarrow \psi(3770) \rightarrow D^0 \bar{D^0}
\end{displaymath}
produces an eigenstate of $CP+$, in the first step, since the $\psi$(3770) has
$J^{PC}$ equal to $1^{--}$. Now consider the case where both the $D^0$ and
the $\bar{D^0}$ decay into $CP$ eigenstates. Then the decays
\begin{displaymath}
\psi(3770) \rightarrow f^i_+ f^j_+ ~~or~~ f^i_- f^j_-
\end{displaymath}
are forbidden, where $f_+$ denotes a $CP+$ eigenstate and $f_-$ denotes a $CP-$
eigenstate. This is because
\begin{displaymath}
CP(f^i_{\pm} ~f^j_{\pm}) = (-1)^\ell = -1
\end{displaymath}
for the $\ell = 1 ~\psi$(3770)
\par
Hence, if a final state such as ($K^+K^-$)($\pi^+\pi^-$) is observed, one
immediately has evidence of $CP$ violation. Moreover, all $CP+$ and $CP-$ 
eigenstates
can be summed over for this measurement.
This measurement can also be performed at higher energies where the final
state $D^{*0} \bar{D^{*0}}$ is produced. When either $D^*$ decays into a
$\pi^0$ and a $D^0$, the situation is the same as above. When the decay is 
$D^{*0} \rightarrow \gamma D^0$ the $CP$ parity is changed by a multiplicative
factor of -1 and all decays $f^i_+ f^j_-$ violate $CP$ \cite{bib:CP}

\subsubsection{Rare Charm Decays}
Rare decays of charmed mesons and baryons provide ``background-free''
probes of new physics effects. In the framework of the Standard Model (SM)
these processes occur only at one loop level. SM predicts vanishingly small
branching ratios for these processes because of the absence in the SM of the
super-heavy bottom-type quark supplemented by almost perfect GIM cancellation
between the contributions of strange and down quarks. This is very different
from the familiar case of bottom quark decays where the top quark contribution
dominates the decay amplitude. It also makes the SM predictions for these
transitions very uncertain, as the pertubative GIM cancellation mechanism
is not effective for soft, long-distance contributions. In addition, in many
cases annihilation topologies also give sizable contribution. At the end, any
anomalous enhancement of a given branching ratio would have to be compared
to the (dominant long-distance) SM amplitude. Fortunately, several
model-dependent estimates exist indicating that the SM predictions for these
processes are still far below current experimental sensitivities. Some examples
are given in \cite{bib:rare1,bib:rare2}. From there is also
follows that experiments which can measure rare $D$ decay branching ratios at 
the level of $10^{-6}$, such as CLEO-c, will start to confront models of new 
physics in an interesting way.

\section{Summary}
The high-precision charm and quarkonium data will permit a broad suite of
studies of weak and strong interaction physics. In the threshold charm sector
measurements are uniquely clean and make possible the unambigous determinations
of physical quantities discussed above. CLEO-c will utilize a variety of tools,
namely $J/\psi$ radiative decays, two-photon collisions (using almost real, as
well as highly virtual space-like photons), deep inelastic Coulomb scattering
and continuum production via $e^+ e^-$ annihilation to obtain significant new
information on the spectrum of hadrons, both normal and exotic, and their decay
channels. A quantitative improvement can be expected not only from the large
accumulated statistics, but also from combining the results obtained using all
these tools together with the results from the $\Upsilon$ resonance runs. The
significance of this is better sensitivity, reduced systematics and a better
chance to obtain a coherent picture of the hadron sector.

\section{Acknowledgments}
I am delighted to acknowledge the invaluable contributions of many individuals
to the development of the CLEO-c and CESR-c program and the outstanding
contributions of my CLEO colleagues over the life of the experiment. The
experimental aspects of this program are based on their effort and experience.

\bibliographystyle{aipproc}   % if natbib is available

\bibliography{hadron_2003}

\IfFileExists{\jobname.bbl}{}
 {\typeout{}
  \typeout{******************************************}
  \typeout{** Please run "bibtex \jobname" to optain}
  \typeout{** the bibliography and then re-run LaTeX}
  \typeout{** twice to fix the references!}
  \typeout{******************************************}
  \typeout{}
 }

\end{document}